\documentclass[%
 aip,
 amsmath,amssymb,
 reprint,%
]{revtex4-1}

\usepackage{graphicx}
\usepackage{dcolumn}
\usepackage{bm}

\usepackage[utf8]{inputenc}
\usepackage[T1]{fontenc}
\usepackage{mathptmx}
\usepackage{etoolbox}

\usepackage{
    lineno,
    textcomp, 
    float,
    mathrsfs
    }

\newcommand{\mycomment}[1]{}
\usepackage{booktabs}
\usepackage{hhline} 
\usepackage{tabularray}

\newcommand{\vect}[1]{\mathbf{#1}}

\newcommand{\maxwell}[0]{\text{s.t. }\ \nabla\times\frac{1}{\mu}\nabla\times\vect{E}-\omega_m^2\vect{\varepsilon}(\rho)\vect{E}=\ -j\omega_m\vect{J}}

\usepackage{upgreek}

\usepackage{xcolor}

\makeatletter
\def\@email#1#2{%
 \endgroup
 \patchcmd{\titleblock@produce}
  {\frontmatter@RRAPformat}
  {\frontmatter@RRAPformat{\produce@RRAP{*#1\href{mailto:#2}{#2}}}\frontmatter@RRAPformat}
  {}{}
}%
\makeatother
\begin{document}

\preprint{AIP/123-QED}


\title[Seeded Topology Optimization for Commercial Foundry Integrated Photonics]{Seeded Topology Optimization for Commercial Foundry Integrated Photonics}
\author{Jacob M. Hiesener}
\author{C. Alex Kaylor}
\author{Joshua J. Wong}
\author{Prankush Agarwal}
\author{Stephen E. Ralph}%
 \email{stephen.ralph@ece.gatech.edu.}
\affiliation{ 
Georgia Institute of Technology, Atlanta, GA
}%

\date{\today}

\begin{abstract}
We present a seeded topology optimization methodology for integrated photonic devices fabricated on foundry platforms that yields improved performance compared to traditional topology optimization. We employ blurring filters and a design rule check correction algorithm to more readily meet fabrication constraints, resulting in devices with fewer artifacts and improved correlation between simulation and measurements. A statistical study is performed on a 2D modal multiplexer, revealing that 87\% of devices optimized using this strategy conform to foundry constraints, compared to 13\% of devices optimized using traditional TO. We apply seeded topology optimization to an ultra-compact TE modal multiplexer, a TE mode converter, a polarization rotator, and a high-contrast grating reflector. Using this optimization strategy, the measured insertion loss of the TE mode converter was reduced from < 1.50 dB to < 0.64 dB, and the measured TE\textsubscript{1} insertion loss of the TE modal multiplexer was reduced from < 3.95 dB to < 1.38 dB over C-band. This approach enables a two-step inverse design process, merging of physics-informed design strategies with inverse design, and ensures strict compliance with foundry constraints throughout optimization.
\end{abstract}

\maketitle

\section{Introduction}
Inverse design is a rapidly evolving method that is used in the design and optimization of integrated photonic devices to create compact structures with record performance that exploit non-intuitive geometries. Density-based topology optimization (TO) is a flexible form of inverse design that allows each voxel of a design (the design parameters) to continuously evolve between two or more materials towards an optimal device topology based on user defined constraints\cite{RN60,RN177,RN174,RN176,RN307}. In density-based TO, two or more Maxwell simulations are performed each iteration to determine the gradient(s) of one or more user-specified figures of merit (FOM) with respect to all design parameters using the adjoint variable method \cite{RN185,RN60}. The gradients for each FOM are linearly combined to update the design region toward a locally optimum topology.

\par

Devices designed for fabrication at a commercial foundry are readily manufactured at high volume after a successful validation process. This ability to scale is achieved by the requirement that devices conform to stringent design rules checks (DRC) to ensure accurate fabrication with high yield \cite{RN182}. We can exploit the gradient descent method used in TO by generating gradients that iterate the design to a condition where design rules are met. In our traditional TO pipeline, we have previously implemented algorithms that calculate gradients to meet geometric linewidth constraints (GLC) and area constraints (AC), which are then linearly combined with the FOM gradient each iteration \cite{RN58}. 

\par

In traditional TO methodologies, the design parameters are initially gray-scale, i.e., allowed to take on any material value between the two or more materials available on that layer in the material stack. However, to accurately detect DRC violations, the device must be sufficiently binary (i.e., every voxel is close to 0 (void) or 1 (solid)), making it challenging to effectively apply DRC constraints. When binarizing the design parameters, a high-performing local optimum in the gray-scale phase of the optimization may not translate to a high-performing local optimum binary topology. Gradient-based optimizers are known to get trapped in local minima valleys or saddle points, which limits the performance achievable through inverse design \cite{RN259,RN187,RN261}. This effect is amplified when fabrication-based constraints are included in the optimization, as conflicting objectives and constraints (FOM vs. AC/GLC) may cause the optimizer to stall when evolving the geometry to satisfy DRC results in a drop in performance \cite{RN62,RN147}. Hence, there is a need for a modified topology optimization algorithm that maintains device fabricability while exploring the local design space around a functional seed geometry.

\par

In this work, we present a seeded TO methodology in which a known functional device geometry, the seed, is iteratively processed and optimized using density-based TO. We demonstrate improved performance compared to traditional TO or conventionally designed seed devices, all designed for a commercial foundry. A limited blurring filter is applied to enable perturbation of the topology such that the design space around the seed can be explored. We develop a DRC correction algorithm that is catered to the seeded TO process, which efficiently resolves foundry constraints while allowing a nested TO stage to improve device performance. 

\par

While traditional TO has the potential to discover completely novel geometries, seeded TO focuses efforts to create strictly fabricable devices. We note that the seed can originate from physics-based conventional design strategies. One example of this is the seeded optimization of high-contrast gratings (HCGs) originally designed using parameter optimization \cite{RN159,RN146}. Shape optimization is a similar inverse design methodology in which a user defines a boundary that is adjusted throughout optimization to maximize performance \cite{RN206}. Previous work has explored seeding shape optimization with a physics-informed initial structure for multimode interferometer or metasurface design \cite{RN179,RN306}. While shape optimization restricts the device topology, allowing for strict enforcement of DRC, our design methodology enables changes in the device topology (i.e., elimination and creation of holes/islands) and includes a DRC correction algorithm that provides more control over minimum linewidths and areas throughout the topology. We apply seeded TO to foundry-fabricated inverse-designed devices and present an improved design pipeline that utilizes the coarse global optimization features of density-based TO followed by the fine, local optimization capability of seeded TO.

\section{Traditional Topology Optimization Overview}

Density-based TO parameterizes the design region such that each voxel can vary between "solid" (high index material) and "void" (low index material) between the two or more materials available on that layer. Typically, every voxel in the design region is initialized to 0.5 or random noise between 0 and 1, meaning the permittivity is between the solid (1) and the void (0), so each voxel can evolve to either. The user specifies the design region lateral dimensions, the location and size of all optical input and output ports (typically waveguides butt-coupling to a fiber is another example), sources and monitors, and formulates one or more FOMs to minimize. Our unique solver is a hybrid time/frequency-domain adjoint-variable method that readily enables solutions across a wide spectrum via the open-source finite-difference time-domain (FDTD) Maxwell solver MEEP \cite{RN165}. This method allows for the inclusion of multiple FOMs for a single device and enables constraint-based TO \cite{RN58}. We formulate TO as a multi-objective minimization problem with $N$ objective functions $(f_1,...,f_N)$ subject to Maxwell's equations at $M$ frequency points, bounds for the design parameters $(\rho)$, and $K$ constraint functions $(g_1,...,g_K)$:

\begin{equation}\label{eq:obj}
    \begin{matrix}
    \min_{\rho}\left[\displaystyle\sum\limits_{n=1}^N\bar{g}_n\big(f_n(\vect{E}),\vect{q}_n\big)+\sum\limits_{k=1}^K\bar{g}_{N+k}\big(g_k(\rho),\vect{q}_{N+k}\big)\right] \\ n\in\left\{1,...,N\right\},\ k\in\{1,...,K\}\\
        \maxwell \\ m\in\left\{1,2,...,M\right\}\\
        0\leq\rho\leq1&\
    \end{matrix}
\end{equation}

where $\bar{g}_i,\ i\in\{1,...,N+K\}$ are differentiable spline-based scaling functions applied to both the objectives and the constraints that are generated using a user defined list of bounds referred to as physical programming bounds ($\vect{q}$) \cite{RN188,RN204}. The returns of each objective/constraint are mapped such that they are optimized on a unified scale, giving the designer more control over the effect of the objective functions and the constraints throughout optimization.

\par

The adjoint variable method is implemented using the in-built solver in MEEP to compute the gradient of the FOM with respect to the design parameters \cite{RN60,RN165}. The gradient is then backpropagated through the parameterization using a vector-Jacobian product implemented via the open-source software package Autograd \cite{autograd}. The latent design variables are optimized using the globally convergent method of moving asymptotes (GCMMA) provided by the open-source nonlinear optimization package NLopt \cite{nlopt}. The GCMMA optimizer produces a sequence of iteration points that are guaranteed to converge to a set of Karush-Kuhn-Tucker points; however, the optimizer may converge to a local minimum with poor performance due to conflicting objectives or constraints \cite{RN255}. A different optimizer may reduce the chance that the optimization gets trapped in a local minimum; however, the GCMMA optimizer allows for the large number of design parameters and constraints required for commercial foundry applications. A comparison of common optimizers used in TO is given in Supp. A.

\par

Though the lithography resolution may limit maximum device performance, traditional TO has been successfully demonstrated to produce ultra-compact, high-performing devices in the constrained design space required for commercial foundries \cite{RN187,RN58,RN180,RN75,RN43,RN44}. The constraints applied to optimizations in this work are GLC (minimum linewidth and linespacing) and AC (minimum area and enclosed area), constraints imposed by the commercial foundry. The gradient for GLC is generated using a chosen set of erosions and dilations that identify inflection regions which violate the minimum linewidth or linespacing, but can also be generated using morphological transforms \cite{RN186,RN187,RN58,RN180,RN181}. AC includes the minimum area and enclosed area, which are implemented using an indicator function that identifies the violating areas to generate a gradient that encourages the holes/islands to dilate or erode. The minimum radius of curvature is another DRC constraint; however, the curvature is implicitly enforced by the GLC implementation and the conic filter ($w$) used to map the design parameters \cite{RN58}. See Supp. B for a mathematical description of the GLC and AC implementations.

\par

The user-set physical programming bounds allow the designer to strategically increase the effect of constraints on the gradient, ensuring that the final device satisfies DRC. These bounds are critical to ensuring the final device is DRC clean but often require hyperparameter tuning through heuristic approaches to mitigate performance loss when AC and GLC physical programming bounds are reduced. Our seeded TO methodology relies less on these bounds, allowing streamlined, efficient optimization that maintains performance with foundry constraints applied.

\subsection{Design Parameter Mapping}

In order to map the latent design parameters ($\rho$) to permittivity values for simulation we use a density-based interpolation scheme. We first filter the design parameters:

\begin{equation}
    \Tilde{\rho} = w * \rho
\end{equation}

where $w$ is a conic filter and $\Tilde{\rho}$ are the filtered design parameters \cite{RN200}. The filtered design parameters are then projected using a differentiable, nonlinear function:

\begin{equation}
    \Bar{\rho} = \frac{\tanh\left( \beta \eta \right) + \tanh\left( \beta \left(\Tilde{\rho} -\eta\right) \right)}{\tanh\left( \beta \eta \right) + \tanh\left( \beta \left(1 -\eta\right) \right)}
\end{equation}

where $\beta$ is a threshold parameter gradually increased throughout the optimization process to binarize the device, $\eta$ is a threshold parameter set to 0.5, and $\Bar{\rho}$ are the mapped design parameters \cite{RN201}. The permittivity is interpolated from the mapped design parameters:

\begin{equation}
    \epsilon_r \left( \Bar{\rho} \right) = \epsilon_{min} + \Bar{\rho} \left( \epsilon_{max} - \epsilon_{min} \right)
\end{equation}

where the relative permittivity of each voxel $\epsilon_r$ varies between the void ($\epsilon_{min}$) and solid permittivity ($\epsilon_{max}$). This linear interpolation scheme works well for silicon photonic devices; however, nonlinear interpolation schemes may be more suitable for other design problems \cite{RN202}.

\section{Seeded Topology Optimization}

\begin{figure*}
    \centering
    \includegraphics[width=6.69in]{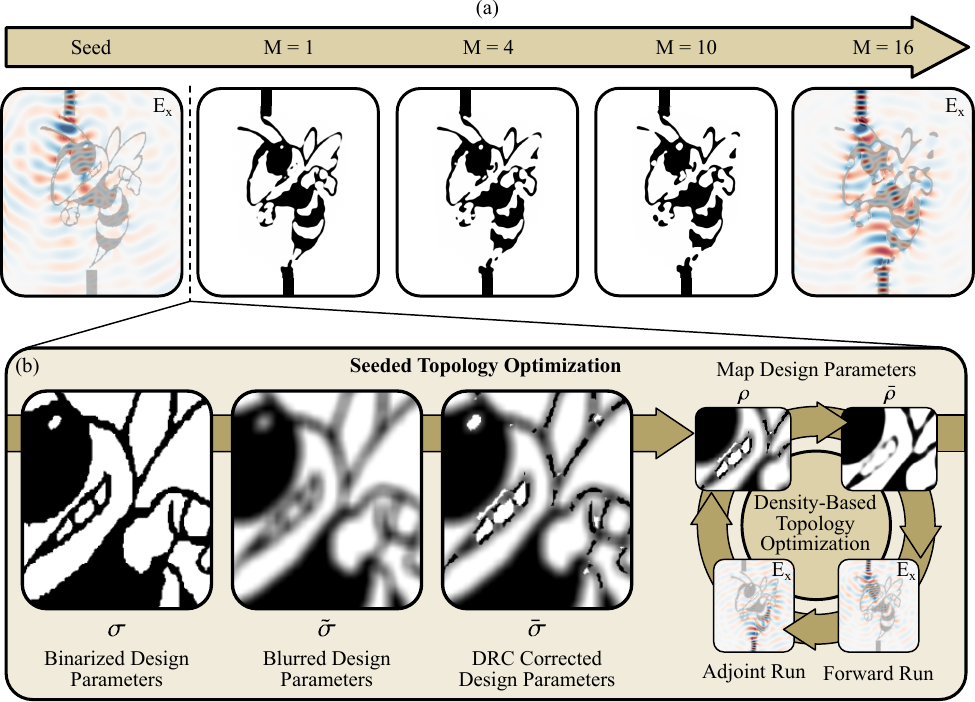} 
    \caption{ Overview of the seeded TO process applied to a seed with poor initial performance and significant DRC violations. (a) 16 iterations (M) of seeded TO are applied to optimize transmission to an output waveguide. (b) Overview of one iteration of seeded TO. First, a blurring filter is applied to the binarized design parameters ($\sigma$) to allow the edges of the device to be perturbed during topology optimization. Second, a DRC correction algorithm is applied to the blurred design parameters ($\Tilde{\sigma}$), adding material in areas violating minimum linewidth/area and removing material in areas violating minimum line-spacing/enclosed area. Finally, design parameters of TO ($\rho$) are set to the DRC corrected design parameters ($\Bar{\sigma}$) and multiple iterations of nested TO are applied to the device. }
\label{fig_sto_overview}
\end{figure*}

The goal of seeded TO is to take a known functional device geometry and improve performance based on a user-specified FOM while maintaining device fabricability. This requires careful processing of the seeded design parameters so the topology can be effectively enhanced using density-based TO. The "seeded" aspect of this methodology refers to initializing the design parameters of seeded TO to a functional device topology. As an extreme example, we use seeded TO to optimize a flawed seed with many DRC violations, achieving a fabricable, semi-functional device (Fig. \ref{fig_sto_overview}a). The DRC violations make this seed ideal for demonstrating how seeded TO can improve device performance while achieving and maintaining DRC clean status. The test cases in this work use functional seeds that comply with DRC and have good initial performance to demonstrate the ability of seeded TO in finding superior local optima.

\par

At the start of each iteration of seeded TO, the mapped design parameters ($\Bar{\rho}$) are first fully binarized:

\begin{equation}
    \sigma \left( \textbf{r} \right) = 
    \begin{cases} 
      0, & \Bar{\rho}\left( \textbf{r} \right) \leq 0.5 \\
      1, & \Bar{\rho}\left( \textbf{r} \right) > 0.5 
   \end{cases}
\end{equation}

where $\sigma$ is the binarized design parameters and $\textbf{r}$ is the position vector of a voxel (Fig. \ref{fig_sto_overview}b). The variable $\sigma$ is used to demarcate the seeded TO design parameters as they are processed differently than the design parameters ($\rho$) used in traditional TO. This binarization stage gives the designer control over the level of blur applied to the design parameters, as any blur introduced via TO or previous seeded TO iterations is removed. The blurred design parameters ($\Tilde{\sigma}$) are then computed:

\begin{equation}
    \Tilde{\sigma} = w_b * \sigma
\end{equation}

where $w_b$ is the blurring filter. This filter is distinct from the conic filter ($w$) used described in sec. 2.1, which enables a smoothed projection of the design parameters to the simulated permittivities, whereas the blurring filter enables perturbation of the topology by the nested TO stage of seeded TO. This blurring stage promotes improvement via TO, since TO can more effectively perturb gray-scale design parameters compared to binary design parameters. The blurring filter is typically set to a box averaging filter of a user-specified size; however, Gaussian filters have also been tested and shown to be effective. The DRC correction algorithm is then applied based on indicator functions used to identify locations in the device topology with DRC violations. The GLC violation indicators (minimum linewidth: $I_{lw}$, minimum linespacing: $I_{ls}$) are found using the open-source software imageruler \cite{RN11,imageruler}, while the AC indicators (minimum area: $I_{a}$, minimum enclosed area: $I_{ea}$) are determined using a contour detection algorithm \cite{RN58}. All the indicators are calculated using the binarized design parameters $\sigma$. The DRC corrected design parameters ($\Bar{\sigma}$) are:

\begin{equation}
    \Bar{\sigma} \left( \textbf{r} \right) = 
    \begin{cases} 
      0, & \textbf{r} \in I_{ls}, I_{ea} \\
      1, & \textbf{r} \in I_{lw}, I_{a} \\
      \Tilde{\sigma}, & \text{otherwise}
   \end{cases} 
\end{equation}

This expands or dilates regions in the mapped topology to algorithmically force DRC compliance beyond inclusion of DRC constraints in TO. Fewer violating inflection regions and contours exist after the DRC correction algorithm is applied, reducing the evaluation of the constraints, which in turn reduces the magnitude of the constraint gradients generated in the nested TO stage. The latent design parameters ($\rho$) of TO are set to the filtered and processed design parameters ($\Bar{\sigma}$):

\begin{equation}
    \rho = \Bar{\sigma}
\end{equation}

and 10-20 iterations of TO are performed (referred to as \textit{nested} TO). Sufficient nested TO iterations must be used to ensure the device reaches a new local minimum, but not so many that the performance plateaus. Nested TO is nearly identical to traditional TO, using the same design parameter mapping scheme as described in Section 2.1 with a high threshold parameter ($\beta$) value (at or above the $\beta$ value used in the final iterations of traditional TO when using TO to generate the seed). AC and GLC must be included in nested TO to mitigate any opposition between the DRC correction and nested TO stages of seeded TO. If DRC constraints are not included in nested TO, nested TO may resist the DRC correction step, causing oscillatory behavior that prevents convergence. This seeded TO cycle is repeated until a high-performing, DRC clean device is achieved.

\par

In traditional TO, careful hyperparameter tuning is required to gradually optimize the device towards a binary topology that satisfies constraints. The traditional TO design pipeline can be transformed into a two-step design process, first with an initial coarse optimization using traditional TO, followed by a fine optimization using seeded TO. This two-step design process uses the broad search space of traditional TO and the DRC correction capability of seeded TO for efficient device optimization.

\subsection{2D Example: Modal Multiplexer}

\begin{figure*}
    \centering
    \includegraphics[width=6.69in]{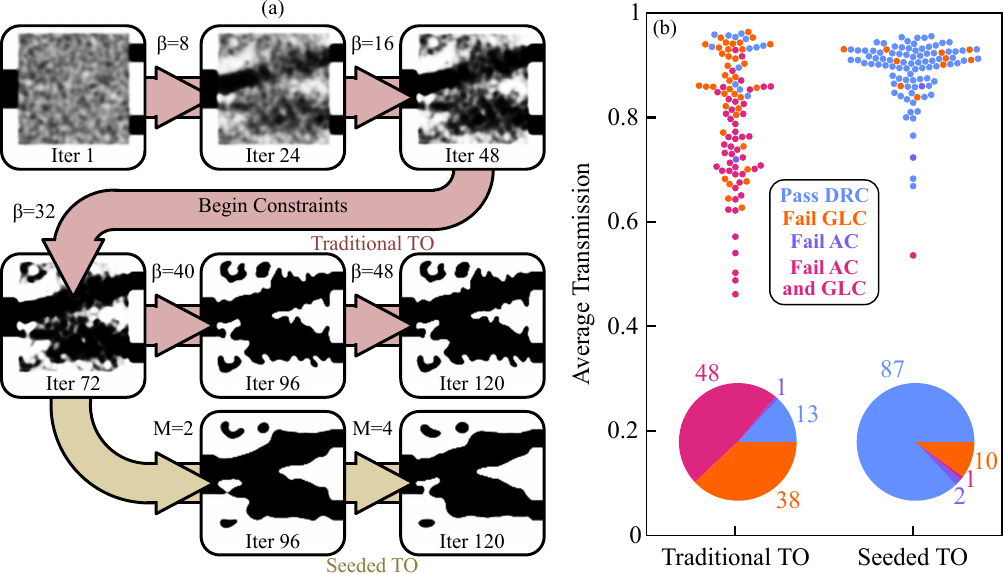} 
    \caption{(a) Optimization process flow for a modal multiplexer using either traditional TO (red) or seeded TO (gold). The design parameters for each run are initialized to randomly generated noise. The threshold parameter ($\beta$) is gradually increased every 24 iterations, and constraints are optimized for starting on iteration 48. 100 independent runs were performed for both traditional TO and seeded TO. (b) Standard swarm plot of the average transmission of both modes over the optimization band for all 100 devices optimized using each method. Pie charts indicate the number of devices that pass DRC (blue), fail geometric linewidth constraints (GLC, orange), fail area constraints (AC, purple), and fail both constraints (red).}
\label{fig_to_v_sto_stat_study}
\end{figure*}

To demonstrate the utility of seeded TO, we perform a statistical study on a modal multiplexer (MMUX), where we design using both a traditional TO and a seeded TO pipeline. We begin by initializing each voxel of the design parameters to a random number from a uniform distribution over the interval $(0,1)$ and optimizing solely for the user-defined FOM while gradually increasing the threshold parameter ($\beta$) to binarize the device. Once the device is sufficiently binary, foundry constraints are added to the optimization. In the traditional TO pipeline, the area and linewidth constraints are increasingly enforced by reducing the physical programming bounds while increasing the threshold parameter ($\beta$) to binarize the device. Reducing the physical programming bounds scales the magnitude of the constraint functions compared to the objective functions, encouraging the optimizer to prioritize minimizing the constraint evaluations. In the seeded TO pipeline, seeded TO is applied to the device shortly after constraints are added to the optimization.

\par

100 MMUX optimizations were performed using 2D FDTD simulations for each design pipeline (traditional TO and seeded TO) (Fig. \ref{fig_to_v_sto_stat_study}a). The design parameters of each optimization were initialized with distinct randomly generated noise; however, TO and seeded TO were performed using the same set of hyperparameters for all runs. This MMUX converts the lowest two TE modes in a multimode waveguide to the fundamental mode of two single-mode waveguides (Sec. 4.1). This optimization problem tends to generate a hole near the multimode waveguide interface which is limited by the minimum area and linespacing constraints, therefore it is an excellent test of the capability of seeded TO in strictly enforcing DRC. Typically, when performing traditional TO, the designer updates the hyperparameters dynamically throughout the optimization to efficiently resolve foundry constraint violations; however, this is challenging when optimizing for multiple FOMs (both modes) due to counteracting gradients. Seeded TO uses both the DRC correction algorithm and nested TO constraint optimization to resolve DRC violations; therefore, it requires minimal hyperparameter tuning to achieve a working device.

\par

The resulting devices were binarized, checked for compliance with GLC and AC, and simulated (Fig. \ref{fig_to_v_sto_stat_study}b). 68 out of 100 seeded TO optimizations yielded devices with superior performance than their traditional TO counterparts. Additionally, 87 of the seeded TO devices conformed to all DRC constraints, while only 13 of the traditional TO devices passed DRC.

\par

Many of the traditional TO devices failed only AC or GLC; these topologies could be adjusted manually after optimization to conform to DRC with minimal performance degradation \cite{RN147}. The highest-performing DRC clean devices optimized using traditional TO achieve the same performance as the highest-performing seeded TO devices, indicating that the selected hyperparameters are effective for these runs, but not for most other traditional TO runs. A majority of the seeded TO devices have above 80\% transmission while maintaining DRC compliance, indicating that the seeded TO process is less sensitive to hyperparameters compared to traditional TO. Though both may satisfy foundry constraints, the seeded TO devices are more suitable for fabrication compared to the traditional TO devices due to the smooth topologies that are generated in the optimization process (Sec. 4).

\subsection{Blurring Filter Study}

\begin{figure}
    \centering
    \includegraphics[width=3.218in]{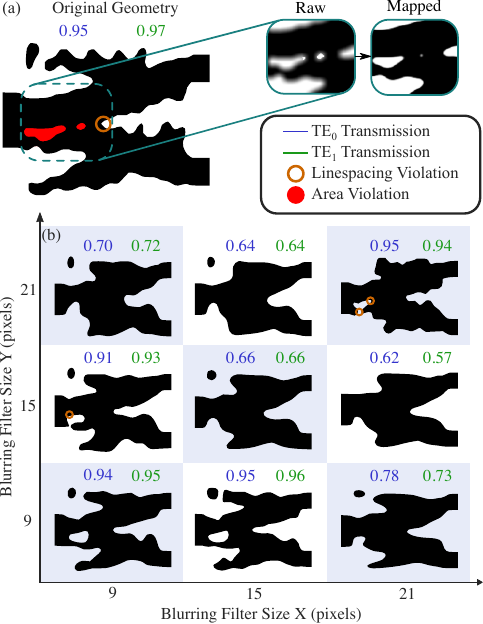} 
    \caption{ (a) Traditional TO was used to design a $2.4$ $\upmu$m $\times$ $2.4$ $\upmu$m modal multiplexer in a 2D simulation environment which contains DRC violations. In seeded TO the design parameters are blurred with a DRC correction algorithm applied. Identical to traditional TO, the parameters are passed through a mapping function to return the permittivity values to be used in the Maxwell simulation. (b) Comparison of varying box filter dimensions used for the seeded TO blurring filter, with DRC violations highlighted. }
\label{fig_filter_study}
\end{figure}

The blurring filter used in seeded TO directly enables perturbation of the design parameters, making the size and shape of the filter critical in altering the overall device performance. Other filter shapes that convert a binary geometry to gray-scale are compatible with seeded TO, here we demonstrate the performance of the box filter with a user-set X and Y size.

\par

The size of the blurring filter significantly impacts the effectiveness of the seeded TO optimization. A small filter has limited ability to perturb the design parameters and therefore limits the design search space to near the seed device. If the filter is too large, small but critical features of the device are obscured and not necessarily recovered in subsequent optimization. An ideal filter size allows for the maximum design region perturbation without eliminating functional features. The size and shape of the optimal filter may vary depending on device and material platform, and will require tuning; however, a practical starting size is one quarter of the single-mode waveguide width for the material platform being used. Since there are only two parameters defining the blurring filter (one if a symmetric filter is used), a parameter sweep can be performed to determine the optimum blurring filter size. Naively, it may appear that the filter size should be on the order of the minimum feature size from the foundry, but in reality, the dimensions are significantly more dependent on the material platform being optimized. For example, silicon nitride designs are longer and have larger features than corresponding silicon designs therefore a larger filter size is required to perturb the device topology \cite{RN235}; however, silicon devices from two foundries with different DRC constraints would have similar optimal filter dimensions as the feature sizes are generally the same.

\par

To investigate the effect of filter size on device performance, a blurring filter study was performed on an ultra-compact MMUX designed using traditional TO with 2D FDTD simulations (Fig. \ref{fig_filter_study}). X and Y filter sizes of 9, 15, and 21 pixels (corresponding to 112.5, 187.5, and 262.5 nm at 80 pixel/$\upmu$m resolution) were selected for the filter study. The MMUX optimized with the 9-pixel $\times$ 9-pixel moving average blurring filter has excellent performance; however, the asymmetric 15-pixel $\times$ 9-pixel filter exhibited slightly better performance. This asymmetric filter has increased blur along the X-axis, which likely elongates the hole at the multimode-waveguide interface without eliminating it, enhancing the TE\textsubscript{1} conversion efficiency. Other size filters had either reduced performance or DRC errors due to the elimination of critical features or minimal design parameter perturbation, respectively. Though the choice of filter size is critical to seeded TO performance, there are no other additional hyperparameters introduced to the optimization, enabling seamless transition between traditional TO and seeded TO. The original hyperparameters, such as the physical programming bounds and thresholding parameters, have less effect on a seeded TO optimization due to the device topology being binary and requiring less tuning for an effective optimization.

\section{Test Cases}

To demonstrate seeded TO we optimize an ultra-compact MMUX \cite{RN147}, a mode converter \cite{RN43}, and a polarization rotator \cite{RN75}, each initially optimized using traditional TO.  We also demonstrate the design of a high contrast grating (HCG) reflector initially optimized using a parameter optimization (PO) method \cite{RN159}. All these devices are designed using 3D FDTD simulations with 40 pixel/$\upmu$m simulation resolution (80 pixel/$\upmu$m design parameter resolution) for the GlobalFoundries silicon photonics process (Fotonix\textsuperscript{TM}). While optimizations using 2D FDTD simulations are sufficient to demonstrate the design process in simulation, 3D FDTD simulations are required to model fabricated devices accurately due to layer thicknesses reducing the effective index of the guiding layer.

\par

Each test case has an objective function depending on the waveguide mode overlap:

\begin{equation}
    a_{m}^\pm = c \int_A\, \left[ \vect{E}^*(r) \times \vect{H}^\pm_m + \vect{E}^\pm_m(r) \times \vect{H}^* \right] \cdot \vect{\hat{n}} \, dA
\end{equation}

where $\alpha^\pm_m$ is the overlap coefficient of the $m^{th}$ mode for forward ($+$) and backward ($-$) directions, $\vect{E}(r)$ and $\vect{H}(r)$ are the Fourier-transformed total fields, $\vect{E}^\pm_m(r)$ and $\vect{H}^\pm_m(r)$ are the mode profiles for the forward and backward propagating modes, and $c$ is the normalization constant \cite{RN58}. The normalization constant is chosen such that:

\begin{equation}
    |\alpha^\pm_m|^2 = \frac{P}{P_{in}}
\end{equation}

where $P$ is the total power propagating in that particular mode, which is normalized to the input power ($P_{in}$), ensuring the maximum value of $|\alpha^\pm_m|^2$ is 1. 

On-chip measured test structures were all created using Fotonix\textsuperscript{TM} PDK grating couplers for optical input and output. A standard fiber array setup was used to couple to the test structures. Spectrum plots were measured using either the LUNA optical vector analyzer (OVA) 5100 or a Keysight 8164B tunable laser source with a Koheron PD10R photodiode. The open-source photonic integrated circuit testing software LabExT was used to automate measurements \cite{RN203}. The grating coupler insertion loss was calibrated out by measuring direct grating-to-grating test structures. Multiple chips across 2 wafer samples were measured for each device.

\subsection{Modal Multiplexer} \label{MMUX_sec}

\begin{figure}
    \centering
    \includegraphics[width=3.202in]{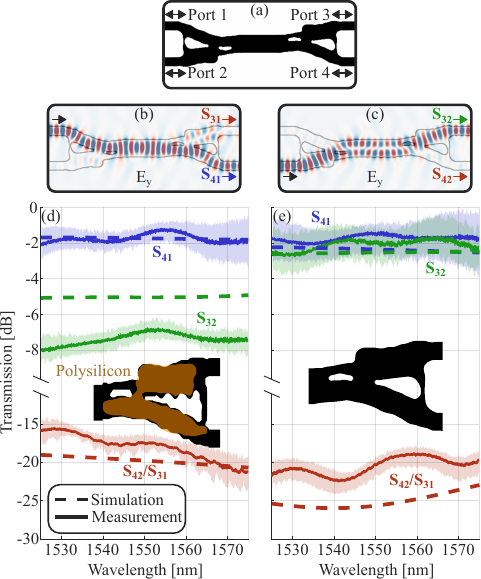} 
    \caption{ (a) Back-to-back measurement test structure for the modal multiplexer with labeled ports. (b) Field plot of the first measurement condition with light input through port 1, transmission measured through port 4, and crosstalk measured through port 3. (c) Field plot of the second measurement condition with light input through port 2, transmission measured through port 3, and crosstalk measured through port 4. The simulated and measured S-parameter spectra are plotted for the traditional TO (d) and seeded TO (e) variants. The dark lines depict the mean values; the light band depicts the worst- and best-case performance across all measured chips. }
\label{fig_2x1cMux_data}
\end{figure}

The use of higher-order optical modes in integrated photonics has many applications, including photonic computation, high extinction/low loss switching systems, and high data-rate communications using mode-division multiplexing \cite{RN190,RN191,RN175}. In these systems, a MMUX or mode converter is required to couple between modes. We use seeded TO to improve an ultra-compact ($3$ $\upmu$m $\times$ $3$ $\upmu$m) MMUX designed with traditional TO (Fig. \ref{fig_2x1cMux_data}a) \cite{RN147}. As described in sec. 3.1, this device converts the fundamental mode of two separate waveguides to the TE\textsubscript{0} and TE\textsubscript{1} modes in a multimode waveguide (Fig. \ref{fig_2x1cMux_data}b,c) \cite{RN180}.

\par

The original design uses both the silicon and polysilicon layers offered by the Fotonix\textsuperscript{TM} process to maximize the number of design parameters available in a compact design region. However, after analysis of experimental results, simulations propounded that the polysilicon layer was not required for an optimal MMUX design \cite{RN62}. Omitting the polysilicon layer from the original MMUX had a negligible effect on simulated device performance, therefore, that layer was removed before performing seeded TO. 

\par

This device requires parallel optimization of both the TE\textsubscript{0} ($f_1(\vect{E})$) and TE\textsubscript{1} ($f_2(\vect{E})$ performance which are defined by the following FOMs:

\begin{equation}
    \begin{matrix}
        f_1 (\vect{E}) = 1 - | \alpha_{0,A}^+ |^2 + b | \alpha_{0,B}^+ |^2  \\ \\ f_2 (\vect{E}) = 1 - | \alpha_{0,B}^+ |^2 + b | \alpha_{0,A}^+ |^2
    \end{matrix}
\end{equation}

$b$ is the extinction coefficient, $\alpha_{0,A}^+$ is the forward propagating fundamental mode coefficient of the single-mode waveguide A (port 1/4), and $\alpha_{0,B}^+$ is the forward propagating fundamental mode coefficient of the single-mode waveguide B (port 2/3) \cite{RN147} (Fig. \ref{fig_2x1cMux_data}a). This is designed to both maximize transmission and minimize the extinction ratio (ER). 

\par

Since no PDK device separates TE modes, modal multiplexers are typically measured using a back-to-back structure shown in Fig. \ref{fig_2x1cMux_data}a \cite{RN185,RN244}. Though the particular mode in the multimode waveguide cannot be determined, the correct modes are assumed to be excited due to mode orthogonality, low crosstalk, and alignment with simulation. Wavelength scans were performed using a Keysight 8164B tunable laser with a Koheron PD10R photodiode. Test structures for both MMUX designs are measured across six separate chips from two wafers and the S-parameters are compared to simulation (Fig. \ref{fig_2x1cMux_data}).

The S\textsubscript{41} transmission encapsulates two passes through the MMUX for the TE\textsubscript{0} channel, while the S\textsubscript{32} transmission encapsulates two passes for the TE\textsubscript{1} channel. The S\textsubscript{42} and S\textsubscript{31} are the same due to reciprocity and capture the crosstalk between the TE\textsubscript{0} and TE\textsubscript{1} channels. 

\par

The traditional TO MMUX required post-TO manual modification to satisfy DRC, which involved expanding the hole near the multimode waveguide to satisfy the minimum area constraint \cite{RN147}. These manual modifications resulted in a significant reduction in the TE\textsubscript{1} transmission in both simulation and measurement for the traditional TO MMUX (Fig. \ref{fig_2x1cMux_data}d). In contrast, seeded TO significantly improved the TE\textsubscript{1} transmission (S\textsubscript{32}) while maintaining the TE\textsubscript{0} transmission (S\textsubscript{41}) (Fig. \ref{fig_2x1cMux_data}e). We note that the FOMs $f_1$ (optimizing TE\textsubscript{0} transmission) and $f_2$ (optimizing TE\textsubscript{1} transmission) were evenly weighed; therefore, the optimizer focused on minimizing the highest FOM in each iteration ($f_2$ for most of the seeded TO run). Through strict enforcement of DRC constraints via our DRC correction algorithm, the seeded TO topology satisfies DRC without manual modifications to the geometry after TO.

\par

\begin{figure*}
    \centering
    \includegraphics[width=5.35in]{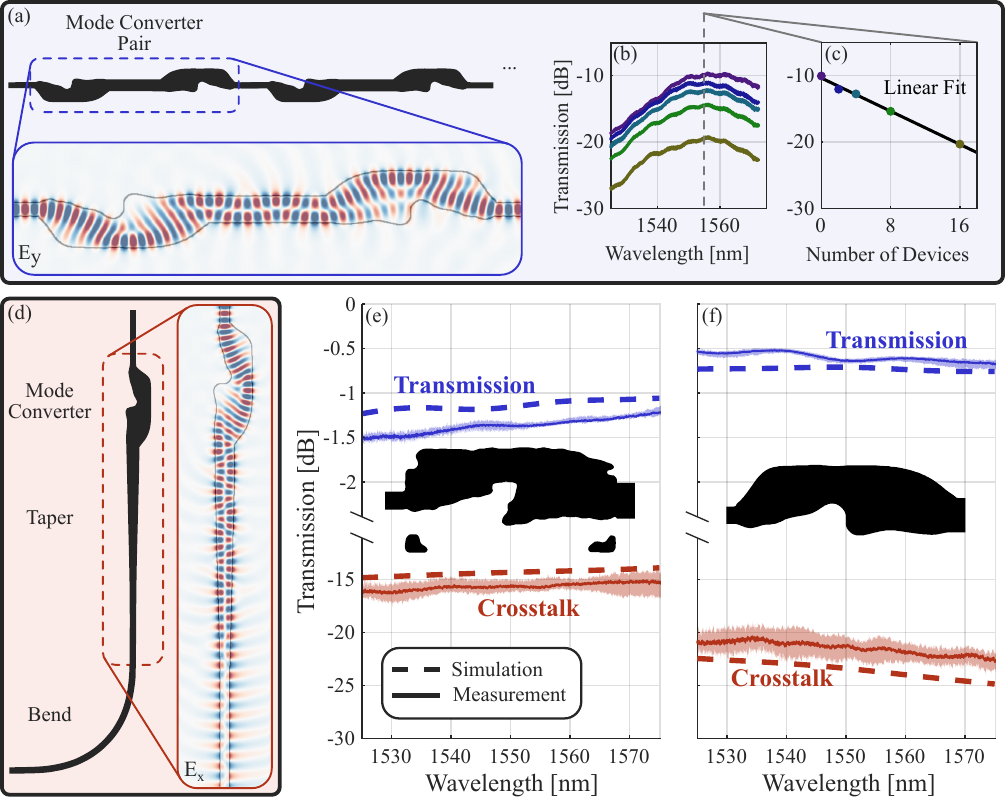} 
    \caption{ (a) The transmission of the TE\textsubscript{0} to TE\textsubscript{1} mode converter is measured using test structures with a varying number of cascaded mode converter pairs. The transmission spectrum is measured for each test structure (b) and a linear fit is applied at each wavelength point (c) to find the loss per device (slope). (d) The crosstalk of the mode converter is measured by using a taper and bend to remove any power in higher-order modes after the mode converter. The simulated and measured transmission and crosstalk spectra are plotted for the traditional TO (e) and seeded TO (f) variants. The dark lines depict the mean values; the light band depicts the worst- and best-case performance across all measured chips. }
\label{fig_MC_data}
\end{figure*}

After experimenting with different blurring filter dimensions, a rectangular 15-pixel $\times$ 9-pixel asymmetric moving average blurring filter is used in seeded TO for this device with a larger blur along the X-axis. This allows the DRC limited hole to stretch in X while preventing annihilation of the hole via a large Y-axis blur, augmenting the mode conversion efficiency for the TE\textsubscript{1} case while maintaining DRC compliance. The final seeded TO topology is smoother than the traditional TO device without the periodic ripples that appear throughout the structure. The disparity in TE\textsubscript{1} simulated and measured transmission for the traditional TO MMUX may be caused by the lithography process used to fabricate these devices which often smooths structures with sharp curvature and small features resulting in variation of device performance compared to simulation \cite{RN153,RN193,RN194}. The seeded TO MMUX has fewer small, jagged features making it more suitable for the lithography process.

\subsection{Mode Converter}

The mode converter designed in this work converts the fundamental TE\textsubscript{0} mode in a single-mode waveguide to the TE\textsubscript{1} mode of a multimode waveguide in a compact ($6$ $\upmu$m $\times$ $3$ $\upmu$m) footprint. The traditional TO version of this device was designed to demonstrate compact, multimode structures for high power signal routing \cite{RN43}. We apply seeded TO to this device to improve the mode conversion efficiency and reduce the modal crosstalk.

\par

Like the MMUX, a TE mode converter is critical for any multimode system. The original device was optimized with a TE\textsubscript{0} source in the single-mode waveguide and a TE\textsubscript{1} monitor in the output waveguide. The objective function was designed to only maximize TE\textsubscript{1} transmission. For the seeded TO device, to experiment with a modified objective function a crosstalk term was added to the objective function that maximizes extinction:

\begin{equation}
    f_1 (\vect{E}) = 1 - | \alpha_{1}^+ |^2 + b | \alpha_{0}^+ |^2 , 
\end{equation}

$b$ is the extinction coefficient, $\alpha_{1}^+$ is the forward propagating mode coefficient for the TE\textsubscript{1} mode, and $\alpha_0^+$ is the forward propagating mode coefficient for the TE\textsubscript{0} mode. This ensures the crosstalk of the final device is low while maximizing mode conversion efficiency. Multiple values of the extinction coefficient were tested for this device (including the original case $b = 0$). The performance of each device was similar; however, $b = 10$ was slightly better in simulation of both transmission and crosstalk, and subsequently the only mode converter selected for fabrication. Seeded TO was performed using a square 11-pixel $\times$ 11-pixel moving average blurring filter; optimization runs with larger and asymmetrical kernels were initially attempted, yielding worse performance improvement. 

Test structures were designed to measure both the transmission and crosstalk of each mode converter. Like the MMUX, wavelength scans were performed using a Keysight 8164B tunable laser with a Koheron PD10R photodiode. The transmission measurement uses test structures with 0, 1, 2, 4, and 8 mode converter pairs (Fig. \ref{fig_MC_data}a). This allows us to apply a linear fit to the transmission vs. number of devices curve at each wavelength point; the slope of the linear fit is the transmission through one device (Fig. \ref{fig_MC_data}b,c). This transmission is the total power through the device, including the TE\textsubscript{0} mode; however, the crosstalk measurement reveals that the TE\textsubscript{0} transmitted power is minimal compared to the total transmission. For the crosstalk measurement, we measure a mode converter followed by a taper and a bend (Fig. \ref{fig_MC_data}d). The taper converts the TE\textsubscript{1} mode to a substrate mode that is lost through the bend. Any power in the fundamental mode after the mode converter is sustained through the taper and bend and measured to determine the crosstalk of the mode converter.

\par

Similar to the Modal Multiplexer, the seeded TO mode converter developed much smoother edges than its traditional TO counterpart. The performance of the seeded TO design is also improved due to lower crosstalk and increased transmission in both simulated and measured data (Fig. \ref{fig_MC_data}e,f). The measured transmission of the traditional TO device is roughly 0.4 dB lower than simulation. Like the MMUX, the traditional TO mode converter may also have a mismatch in performance due to lithographic smoothing. However, the seeded TO mode converter has slightly higher measured transmission than simulation. This indicates that features developed through seeded TO are more suitable to the fabrication process.

\subsection{Polarization Rotator}

A polarization rotator converts the fundamental TE mode to the fundamental TM mode in a single-mode waveguide. The original polarization rotator was designed using traditional TO with a size of $8$ $\upmu$m $\times$ $2$ $\upmu$m \cite{RN75}. Polarization control in integrated photonics enables a variety of applications including polarization division multiplexing, medical sensing, and dispersion engineering \cite{RN44,RN196,RN197,RN198}. To convert from TE to TM the direction of the $\Bar{E}$ and $\Bar{H}$ fields needs to rotate by $90^\circ$; this can only be achieved using a structure that breaks z-symmetry, which the polysilicon layer on the Fotonix\textsuperscript{TM} process can be used for, making it crucial for an effective polarization rotator \cite{RN75}. Both the silicon and polysilicon layers are optimized simultaneously throughout the traditional TO process.

\begin{figure}
    \centering
    \includegraphics[width=3.298in]{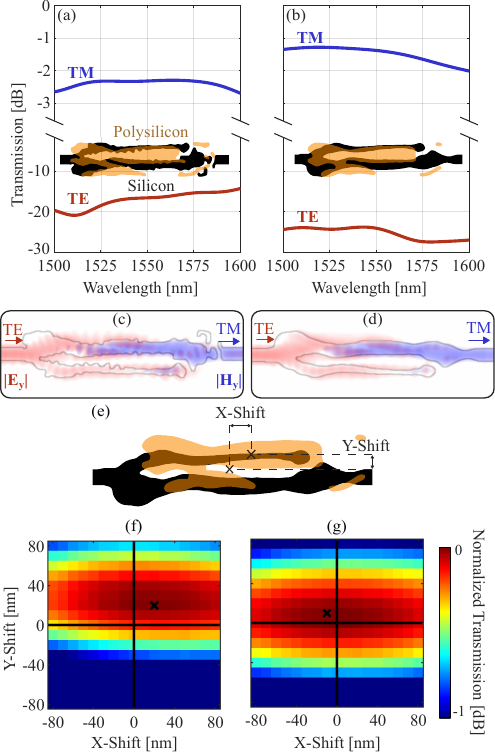} 
    \caption{  Simulations reveal that TM transmission and TE extinction spectra are uniformly improved for the seeded TO (b) compared to the traditional TO  design (a) of a polarization rotator. (c,d) Normalized log-scale field magnitudes show the rotation of the TE mode via the $|E_y|$ field (red) to the TM mode via the $|H_y|$ field (blue) for both devices. (e) A layer misalignment study was performed by shifting the polysilicon layer from the nominal position in both X and Y. Heat maps are generated for the polarization rotator transmission at 1550 nm normalized to the maximum transmission (marked with an "x") for the traditional TO (f) and the seeded TO (g) polarization rotator. }
\label{fig_pol_rotator_data}
\end{figure}

The objective function used to optimize the polarization rotator was designed solely to maximize the TE to TM conversion efficiency:

\begin{equation}
    f_1 (\vect{E}) = 1 - | \alpha_1^+ |^2
\end{equation}

\begin{figure*}
    \centering
    \includegraphics[width=6.69in]{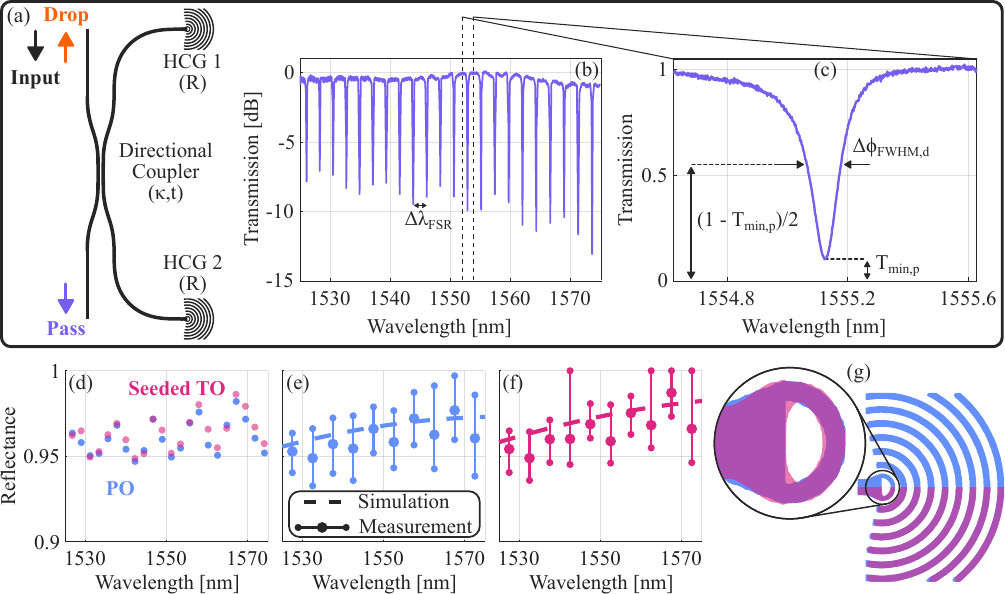}
    \caption{ (a) Resonator structure measured to determine the reflectance of the high contrast grating (HCG). (b,c) Measured pass-port transmission spectrum of the resonator with marked FSR ($\Delta\lambda_{FSR}$), drop-port FWHM ($\Delta\lambda_{FWHM,d}$), and pass-port minimum transmission ($T_{min,p}$). (d) The measured reflectances for both devices from a single chip. The simulated and measured reflectance spectra from all 5 chips are plotted for the parameter optimized (PO) (e) and seeded TO (f) variants. The measured reflectances at each resonance peak are binned in 5 nm intervals; the mean, min, and max of each bin are shown. (g) Comparison of PO and seeded TO device geometries. }
\label{fig_HCG_data}
\end{figure*}

where $\alpha_1^+$ is the first TM mode coefficient propagating forward. The TO polarization rotator is optimized with seeded TO using a square 11-pixel $\times$ 11-pixel moving average blurring filter for both the silicon and polysilicon layers. Simulation results reveal the seeded TO polarization rotator outperforms the traditional TO polarization rotator in both crosstalk and transmission over C-band (Fig. \ref{fig_pol_rotator_data}a,b). Test structures for this device have been included in a future multi-process wafer tapeout.

\par

Like the previous devices, seeded TO smoothed many of the ripples present in the traditional TO polarization rotator. The output waveguide is disconnected from the bulk of the traditional TO device whereas the output waveguide is connected in the seeded TO variant. A shape optimization design methodology would not allow for disconnected features to merge, changing the device topology \cite{RN179,RN206}. The improvement made by seeded TO can be seen in the field plots where the magnitude of the $E_y$ field is significantly reduced in the output waveguide of the seeded TO variant compared to the traditional TO device (Fig. \ref{fig_pol_rotator_data}c,d). 

\par

To investigate the robustness of seeded TO, we performed a layer misalignment study on the polarization rotator (Fig. \ref{fig_pol_rotator_data}e-g). The polysilicon layer was shifted in both X and Y with respect to the nominal position. Both devices are far more sensitive to layer misalignment in the Y-direction due to the small device width. The peak transmission for the traditional TO polarization rotator is shifted 20 nm in X and 20 nm in Y from the nominal position whereas the peak transmission for the seeded TO device is shifted -10 nm in X and 10 nm in Y. The optimal layer alignment for the seeded TO variant is closer to the nominal position than the traditional TO variant, indicating that the local optimization behavior of seeded TO yields a stronger local optimum than traditional TO.

\subsection{High Contrast Grating Reflector}

Parameter optimization (PO) was used to generate a $4.5$ $\upmu$m $\times$ $6$ $\upmu$m HCG to investigate the performance of seeded TO with a seed designed using alternative methods to TO. This HCG reflector is designed to reflect the fundamental mode of a single-mode waveguide with an ultra-compact footprint. These reflectors are fundamental in the design of many integrated photonic systems such as compact filters and integrated lasers \cite{RN195}. The PO HCG consists of apodized concave gratings with a taper to shape the waveguide mode entering the grating region \cite{RN159}. The design of the PO HCG was based on other high contrast gratings, such as grating couplers and circular grating reflectors \cite{RN160}. Seeded TO was then applied to the PO structure using a square 11-pixel $\times$ 11-pixel moving average blurring filter to further improve the reflectance of the grating. 

Mirror symmetry can be applied to this device about the $y = 0$ axis. The FOM used for the optimization of this device is to maximize transmission into the reflected mode:

\begin{equation}
    f_1 (\vect{E}) = 1 - | \alpha_0^- |^2
\end{equation}

$\alpha_0^-$ is the fundamental mode coefficient propagating backwards. To measure this device, a resonator was created using a directional coupler and two HCGs as mirrors (Fig. \ref{fig_HCG_data}a). This device acts as an add/drop ring resonator due to the counter-propagating light in the resonator. The HCG reflectance ($R$) is related to the drop port FWHM ($\Delta\lambda_{FWHM,d}$), the FSR ($\Delta\lambda_{FSR}$), and the directional coupler through-coupling coefficient ($t$) through the following equation \cite{RN153,RN184}:

\begin{equation}
    \frac{\Delta\lambda_{FWHM,d}}{\Delta\lambda_{FSR}} = \frac{2}{\pi}\sin^{-1}\left( \frac{1-Rt^2}{2\sqrt{R}t}\right)
\end{equation}

The derivation for this can be found in Supp. C. Like the previous devices, the resonator pass-port test structure was implemented using the Fotonix\textsuperscript{TM} PDK grating couplers for I/O to be measured using a standard fiber array testing setup. The LUNA OVA 5100 was used for this measurement to ensure sufficient wavelength resolution (1.2 pm for the LUNA OVA compared to 10 pm for the tunable laser sweep). Resonant peaks in the transmission spectrum are identified using a peak-finding algorithm and used to calculate the FSR (Fig. \ref{fig_HCG_data}b). Each peak is isolated and the FWHM is algorithmically calculated to determine the reflectance at each resonance peak (Fig. \ref{fig_HCG_data}c). Test structures are measured across 5 separate chips from 2 wafer samples, the calculated reflectances are binned in 5 nm intervals, and the mean, min, and max of each bin are calculated and plotted along with the simulated reflectance (Fig. \ref{fig_HCG_data}d,e,f).

\par

Seeded TO had limited visible effect on the topology of the HCG, only modifying the hole at the waveguide-grating interface (Fig. \ref{fig_HCG_data}g). While the seeded TO variant required no manual modification of the topology to conform to DRC, the PO variant required modification of the corners of the small semicircle hole to conform to the minimum curvature radius constraint. This feature is critical to achieving maximum performance through shaping of the light entering the grating, but is limited by the minimum enclosed area constraint. Though there is some variance in the measured response, the seeded TO modifications are demonstrated to improve performance in both simulation and measurement. Figure \ref{fig_HCG_data}d shows that the calculated reflectance of the seeded TO variant is typically higher than the PO variant at corresponding resonances. The average measured/simulated reflectance over C-band is 0.963/0.969 for the seeded TO HCG compared to 0.960/0.963 for the PO variant. This validates the effect that the minimal modifications made by seeded TO have on the performance of fabricated devices. The changes made by seeded TO around the waveguide-grating interface are small enough that a designer may assume they will not be resolved in fabrication. However, the measured improvement of the fabricated device demonstrates that small changes like this are critical to device performance. 

\section{Conclusion}

We have demonstrated a seeded TO design methodology that enables effective optimization of commercial foundry-compatible integrated photonic devices, yielding improved performance over traditional TO. The seed is best chosen as a known functional device created via traditional TO, other optimization methods such as parameter or shape optimization, or from physics-based models. Seeded TO relies on a blurring filter chosen to perturb the known structure seeking an optimized design. This new design methodology enables the creation of new, more robust algorithms to ensure the device meets DRC. We illustrated this optimization technique using four different test devices designed for a foundry process.

\par

Seeded TO brings several important benefits to the integrated photonics inverse design community not seen in traditional topology optimization implementations. While traditional TO has the capability to discover non-intuitive device geometries, the optimal topology changes throughout the optimization as the design parameters binarize and DRC constraints are applied. This limits the ability of traditional TO to find a strong optimal topology without careful hyperparameter tuning or a post-TO optimization scheme. In many TO implementations hyperparameter tuning is done heuristically by executing a TO algorithm and evaluating the solution upon completion \cite{RN244,RN245}. Less hyperparameter tuning is required when using seeded TO resulting in fewer reloads throughout the optimization, potentially reducing the computational cost of an optimization. This allows the designer to focus on the other aspects of the design process as hyperparameter tuning is often a tedious, time-consuming process. The computational cost of an optimization can be further reduced by performing traditional TO using low-resolution or 2D FDTD simulations followed by full resolution seeded TO \cite{jmh_cleo_2025}.

\par

Though only DRC fabrication constraints were considered in this work, additional TO constraints and permittivity mapping schemes are compatible with seeded TO. This includes constraints on the etching process for multilayer designs and permittivity projection operations used to optimize devices on platforms with non-vertical sidewalls \cite{RN44,RN204,RN225,RN222,RN232}. A desirable consequence of seeded TO is the elimination of the periodic ripples that are commonly develop through traditional TO and have limited impact on device performance. This reduces the effect lithographic smoothing has on the device topology, resulting in greater alignment between measured and simulated performance.

\par

There remain many opportunities for future work including exploring potential design techniques that can be used to create the seed (e.g. inverse design, shape optimization, physics-defined design, etc.). Designing a seed using a physics-defined topology will allow for the optimization of large structures such as multimode interferometers, spot-size converters, and Bragg grating filters that have traditionally been difficult to inverse design due to simulation complexity. Microcavity design using local density of states is a common design problem in TO which is sensitive to small perturbations and can develop tiny features that prevent fabrication \cite{RN234}, a seeded TO method may assist in improving the fabricability and robustness of these devices. Since seeded TO performs filtering and DRC correction outside of TO, additional constraints such as requiring all features of a device to be connected can easily be incorporated into seeded TO \cite{RN236,RN176}. There is scope to explore additional functionality in seeded TO such as creating algorithms that identify non-essential features of TO structures that can be removed to reduce device footprint. Intelligent implementation of non-gradient based design parameter modification has scope to significantly improve the performance of devices designed using TO.

\section{Supplemental Material}

\subsection{TO Optimizers}

TO typically requires a gradient-based nonlinear optimization method to calculate the design parameter step for each iteration. Two commonly employed methods are the method of moving asymptotes (MMA) or L-BFGS-B \cite{RN255,RN264}. In this work we employ the globally convergent method of moving asymptotes; however, different optimizers have different constraints and produce different optimization trajectories which need to be considered when designing for a specific application.

\par

MMA uses simple functions with penalty terms to approximate the nonlinear function and determine the next step. L-BFGS-B uses the gradient of the objective function and a limited-memory approximation of the hessian derived from previous iterations to determine the descent direction in the presence of bound constraints. Both algorithms are well-suited for large-scale optimization problems with thousands of design variables. However, L-BFGS-B only accepts bound constraints i.e. constraints where the design variable falls between scalar limits. MMA, on the other hand, also permits constraints that are nonlinear functions of the design variables, allowing more freedom in choice of constraints. The commercial foundry integrated photonics design problem introduces many foundry-based design constraints an MMA optimizer can handle well.

\par

L-BFGS-B applied to photonics problems tends to produce optimization paths that are very nearly monotonic since it exploits Hessian information to obtain an accurate update direction \cite{RN186}. Furthermore, due to the update precision L-BFGS-B may require only 100 iterations to stop improving the objective function significantly. MMA takes suboptimal steps that result in optimization paths with spikes \cite{RN261,RN58}. The spikes depend greatly on the specific problem being solved, for example bends contain few while a broadband mirror has many \cite{RN58}. MMA also requires on the order of 200 iterations to produce an optimized design. There are many other optimizers explored in the inverse design community, each with benefits and drawbacks that make no optimizer ideal for all integrated photonic applications.

\subsection{Foundry-Set Constraint Implementations}

In this section we outline our design rule check (DRC) constraint implementation for geometric linewidth constraints (GLC) and area constraints (AC).

\subsubsection{Geometric Linewidth Constraints}

GLC includes the minimum linewidth and linespacing which are the minimum lengthscales of solid and void features that can be accurately fabricated. These constraints are commonly combined and referred to as the minimum feature size; however, many platforms have different values for the minimum linewidth and linespacing, making it useful to separate these constraints. The minimum linewidth constraint ($g_{LW}$) is described by the function

\begin{equation}
    g_{LW} = \frac{1}{n} \sum_{i\in N} I_i^{WL} (\rho_i) \cdot \left[ \text{min} \{ \left( \Tilde{\rho} - \eta_e\right), 0 \} \right]^2
\end{equation}

where $n$ is the number of inflection regions ($N$) identified that violate the minimum linewidth, $\Bar{\rho}$ is the projected design parameters, $I_i^{WL} (\rho_i)$ is the indicator function that identifies each inflection region of the solid phase, and $\eta_e$ is the linewidth threshold parameter \cite{RN58}. The indicator function is defined as

\begin{equation}
    I_{i}^{LW} (\rho) = \Bar{\rho} \cdot \exp{\left( -c \left| \nabla \Tilde{\rho} \right|^2 \right)}
\end{equation}

where $c$ is a dampening term that dictates the strength of the indicator function, this is typically set to $r^4$ where $r$ is the design grid resolution \cite{RN58}. The linewidth threshold parameter is given by

\begin{equation}
    \eta_e = 
    \begin{cases} 
      \frac{1}{4} \left( \frac{l_W}{R} \right)^2 + \frac{1}{2} & \frac{l_W}{R} \in \left[0,1\right] \\
      -\frac{1}{4} \left( \frac{l_W}{R} \right)^2 + \frac{l_W}{R} & \frac{l_W}{R} \in \left[1,2\right] \\
      1 & \frac{l_W}{R} \in \left[2,\infty\right]
   \end{cases}
\end{equation}

where $l_w$ is the minimum linewidth and $R$ is the user-specified radius of the conic filter \cite{RN58}. This allows the user to arbitrarily choose the filter radius without dependence on the foundry constraints. Similarly, the minimum linespacing constraint ($g_{LS}$) is described by

\begin{equation}
    g_{LS} = \frac{1}{n} \sum_{i\in N} I_i^{LS} (\rho_i) \cdot \left[ \text{min} \{ \left(\eta_d - \Tilde{\rho}\right), 0 \} \right]^2
\end{equation}

where the indicator function ($I_i^{LS} (\rho_i)$) which identifies the inflection region of the void phase is given by

\begin{equation}
    I_{i}^{LS} (\rho) = \left(1 - \Bar{\rho}\right) \cdot \exp{\left( -c \left| \nabla \Tilde{\rho} \right|^2 \right)}
\end{equation}

The linespacing threshold parameter ($\eta_d$) is 

\begin{equation}
    \eta_d = 
    \begin{cases} 
      \frac{1}{2} - \frac{1}{4} \left( \frac{l_S}{R} \right)^2 & \frac{l_S}{R} \in \left[0,1\right] \\
      1 + \frac{1}{4} \left( \frac{l_S}{R} \right)^2 - \frac{l_S}{R} & \frac{l_S}{R} \in \left[1,2\right] \\
      0 & \frac{l_S}{R} \in \left[2,\infty\right]
   \end{cases}
\end{equation}

where $l_S$ is the minimum linespacing \cite{RN228}. Using circular filters, the linewidth and linespacing implementations both impose implicit constraints on the corresponding minimum curvature for both solid and void regions. The minimum radius of curvature is given by \cite{RN58}

\begin{equation}
    k_{W,S} = \frac{l_{W,S}}{2}
\end{equation}

\subsubsection{Minimum Area Constraints}

AC includes the minimum area and enclosed area constraints, which dictate the smallest allowable island or hole that can be accurately fabricated. The minimum area constraint function ($g_A$) is defined as

\begin{equation}
    g_A = \sum_{i\in N} \sin\left( \frac{\pi}{A_{min}} A_i \left( \Bar{\rho} , I_A (\Bar{\rho})\right)\right)
\end{equation}

where $N$ are the contours of the topology that violate the minimum area, $A_{min}$ is the minimum area, $A_i \left( \Bar{\rho}, I_A (\Bar{\rho})\right)$ is the area of the $i^{th}$ contour, $\Bar{\rho}$ are the projected design parameters and $I_A (\Bar{\rho})$ is an indicator function that is marks all regions of the topology that contain \textit{islands} with areas below the minimum area constraint. The minimum enclosed area constraint function ($g_{EA}$) is defined as

\begin{equation}
    g_{EA} = \sum_{i\in M} \sin\left( \frac{\pi}{E_{min}} E_i \left( 1 - \Bar{\rho} , I_{EA} (1 - \Bar{\rho})\right)\right)
\end{equation}

where $M$ is the contours of the topology that violate the minimum enclosed area, $E_{min}$ is the minimum enclosed area, $E_i \left( 1-\Bar{\rho}, I_{EA} (1-\Bar{\rho})\right)$ is the area of the $i^{th}$ contour, and $I_{EA}(1-\Bar{\rho})$ is an indicator function that marks all regions of the topology that contain \textit{holes} with areas below the minimum enclosed area constraint. 

\par

The indicators ($I_A$, $I_{EA}$) are determined using an out-of-the box python package to perform the marching-squares algorithm to extract contours from the design parameters ($\Bar{\rho}$, $1 - \Bar{\rho}$) \cite{RN230,RN229}. With the contours identified, the area of each contour ($A_i$, $E_i$) is calculated using another out-of-the box Python function that calculates areas using a discrete summation of all the density values inside the contour, which are identified using morphological dilations \cite{RN231, RN58}. If a particular contour has a smaller area than the minimum area, the filled contour region is dilated by 1 pixel and added to the indicator function.

\par

The constraints are defined as

\begin{equation}
    g_{A,EA} \leq 0
\end{equation}

such that the optimizer drives each constraint to 0 \cite{RN58}. The $\sin$ in the constraint functions enables both erosion and dilation of islands and holes depending on the size of the violating feature. If the area of an island is less than half of the minimum area, it is eroded, whereas if the area is larger than half the minimum area, it is expanded. 

\subsection{High Contrast Grating Reflectance From Resonator Response}

\begin{figure}
    \centering
    \includegraphics[width=3.274in]{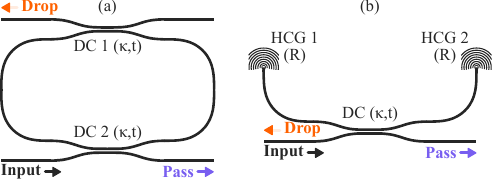} 
    \caption{ (a) Add/drop ring resonator and (b) high contrast grating (HCG) resonator with directional couplers (DC), input ports, through ports, and drop ports labeled.}
\label{fig_HCG_diagram_sup}
\end{figure}

To measure the high contrast grating (HCG), a resonator was created with a directional coupler and 2 HCGs as mirrors (Fig. \ref{fig_HCG_diagram_sup}). This device acts as an add/drop ring resonator due to the counter-propagating light in the resonator. We derive the relation between the FWHM ($\Delta\lambda_{FWHM,d}$), FSR ($\Delta\lambda_{FSR}$), and directional coupler through-coupling coefficient ($t$) starting from the drop-port response:

\begin{equation}
    \frac{E_{drop}}{E_{in}} = \frac{-\kappa^2 A^{\frac{1}{4}} e^{j\phi/2}}{1 - \sqrt{A} t^2 e^{j\phi}}
\end{equation}

where $\kappa = \sqrt{1-t^2}$ is the directional coupler cross-coupling coefficient, $t$ is the directional coupler self-coupling coefficient (assumed real), $A$ is the round-trip optical power attenuation, and $\phi$ is the round-trip optical phase \cite{RN153}. The transmission to the drop port is:

\begin{multline}
    T_d\left( \phi \right) = \frac{I_{drop}}{I_{in}} = \left| \frac{E_{drop}}{E_{in}} \right|^2 = \frac{\kappa^4 \sqrt{A}}{1 - t^2 \sqrt{A} \left( e^{j\phi} + e^{-j\phi} \right) + t^4 A} \\ = \frac{\kappa^4 \sqrt{A}}{1 - 2 t^2 \sqrt{A} \cos\phi + t^4 A}
\end{multline}

The round-trip optical power loss is given by $A = R^2 e^{-\alpha L_{rt}}$ where $R$ is the mirror reflectance and $e^{-\alpha L_{rt}}$ encapsulates the waveguide propagation loss. For this resonator, we assume the loss is all due to the mirror reflectance ($A = R^2$), the transmitted power becomes:

\begin{equation}
    T_d\left( \phi \right) = \frac{R\kappa^4}{1 + R^2t^4 - 2 R t^2 \cos\phi}
\end{equation}

We can do the same calculation for the pass port, the transmission to the pass port is:

\begin{equation}
    T_p\left( \phi \right) = \frac{t^2 + R^2t^2 - 2 R t^2 \cos\phi}{1 + R^2 t^4 - 2 R t^2 \cos\phi}
\end{equation}

\begin{figure}
    \centering
    \includegraphics[width=3.27in]{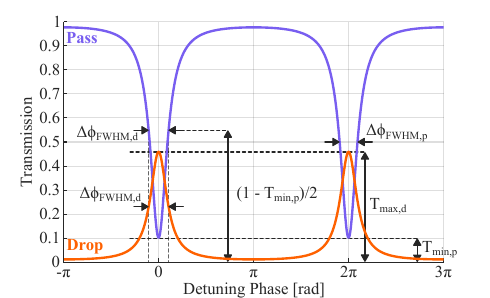} 
    \caption{ Transmission spectrum of the add-drop ring resonator with labeled parameters.}
\label{fig_phase_sup}
\end{figure}

The transmission to the pass and drop port are shown in Fig. \ref{fig_phase_sup}. The maximum transmission at the drop port is:

\begin{equation}
    T_{max,d} = \frac{R \kappa^4}{1 + R^2 t^4- 2 R t^2} = \frac{R\kappa^4 }{\left( 1 - R t^2\right)^2}
\end{equation}

We can rewrite the drop port transmission as a function of the maximum transmission at the drop port:

\begin{multline}
    T_d\left( \phi \right) = \frac{T_{d,max}\left( 1 - R t^2\right)^2}{1 + R^2 t^4 - 2 R t^2 (1-2\sin^2\left(\phi/2\right))} = \\ \frac{T_{d,max}\left( 1 - R t^2\right)^2}{\left( 1 - R t^2\right)^2 + 4Rt^2\sin^2\left(\phi/2\right)}
\end{multline}

We can define the contrast of the resonator $F$:

\begin{equation}
    F = \frac{4Rt^2}{\left( 1 - Rt^2\right)^2}
\end{equation}

\begin{equation}
    T_d\left( \phi \right) = \frac{T_{d,max}}{1 + F\sin^2\left(\phi/2\right)}
\end{equation}

We can equate the drop port power to the maximum value to determine the phase bandwidth ($\Delta\phi_{FWHM,d}$):
  
\begin{multline}
    T_d\left( \Delta\phi_{FWHM,d}/2\right) = \frac{T_{d,max}}{2} \rightarrow \\ \frac{T_{d,max}}{1 + F\sin^2\left(\Delta\phi_{FWHM,d}/4\right)} = \frac{T_{d,max}}{2}
\end{multline}

Simplifying:

\begin{equation}
    \Delta \phi_{FWHM,d} = 4sin^{-1}\left( \frac{1}{\sqrt{F}}\right) = 4sin^{-1}\left( \frac{1-Rt^2}{2\sqrt{R}t}\right)
\end{equation}

The phase bandwidth is related to the FWHM ($\Delta\lambda_{FWHM}$) and the FSR ($\Delta\lambda_{FSR}$) by \cite{RN184}:

\begin{equation}
    \frac{\Delta\lambda_{FWHM}}{\Delta\lambda_{FSR}} = \frac{\Delta\phi_{FWHM}}{2\pi}
\end{equation}

This becomes:

\begin{equation}
    \frac{\Delta\lambda_{FWHM,d}}{\Delta\lambda_{FSR}} = \frac{2}{\pi}\sin^{-1}\left( \frac{1-Rt^2}{2\sqrt{R}t}\right)
\end{equation}

This equation relates the drop-port FWHM (measured at $T_p = (1 + T_{min,p})/2$ (see Fig. \ref{fig_phase_sup})), FSR, and directional coupler self-coupling coefficient to the mirror reflectance. 

\begin{acknowledgments}
This material is based upon work supported in part by the National Science Foundation (NSF) Center ``EPICA” under Grant No.1 2052808, \url{https://epica.research.gatech.edu/}. Any opinions, findings, and conclusions or recommendations expressed in this material are those of the author(s) and do not necessarily reflect the  views of the NSF. JMH, CAK, JJW, PA, and SER were supported by the Georgia Electronic Design Center of the Georgia Institute of Technology. This research was supported in part through research cyberinfrastructure resources and services provided by the Partnership for an Advanced Computing Environment (PACE) at the Georgia Institute of Technology, Atlanta, Georgia, USA \cite{PACE}. The authors would like to thank GlobalFoundries for providing silicon fabrication through Fotonix\textsuperscript{TM} university program.
\end{acknowledgments}

\section*{Data Availability Statement}

The data that support the findings of this study are available from the corresponding author upon reasonable request.

\section*{References}
\nocite{*}
\bibliography{bib}

\end{document}